\documentclass[eqsecnum,aps,prd,nofootinbib]{revtex4}
\usepackage[pdftex]{graphicx}

\def\ben{\begin{equation}}
\def\een{\end{equation}}
\def\bena{\begin{eqnarray}}
\def\eena{\end{eqnarray}}

\usepackage{amsmath}
\usepackage{enumerate}

\newcommand{\eps}{{\mbox{\boldmath $\epsilon$}}}

\newcommand{\T}{{\bf T}}

\begin{document}
\title {Counter-term charges generate bulk symmetries}

\author{Stefan Hollands}

\affiliation{Institut f\" ur Theoretische Physik, U. G\" ottingen, \\
D-37077 G\" ottingen, Germany and\\
Physics Department, UCSB, Santa Barbara, CA 93106, USA\\
{\tt hollands@vulcan2.physics.ucsb.edu}}

\author{Akihiro Ishibashi}

\affiliation{Enrico Fermi Inst., U Chicago, Chicago, IL 60637, USA\\
{\tt akihiro@midway.uchicago.edu}}

\author{Donald Marolf}

\affiliation{Physics Department, UCSB, Santa Barbara, CA 93106, USA\\
{\tt marolf@physics.syr.edu}}

% \date{February, 2005}

\begin{abstract}
We further explore the counter-term subtraction definition of
charges (e.g., energy) for classical gravitating theories in
spacetimes of relevance to gauge/gravity dualities; i.e., in
asymptotically anti-de Sitter spaces and their kin.   In
particular, we show in general that charges defined via the
counter-term subtraction method generate the desired asymptotic
symmetries. As a result, they can differ from any other such charges,
such as those defined by bulk spacetime-covariant
techniques, only by a function of auxiliary non-dynamical
structures such as a choice of conformal frame at infinity
(i.e., a function of the boundary fields alone). Our
argument is based on the Peierls bracket, and in the AdS context
allows us to demonstrate the above result even for asymptotic
symmetries which generate only conformal symmetries of the
boundary (in the chosen conformal frame).
We also generalize the counter-term subtraction construction of charges
to the case in which additional non-vanishing boundary fields are present.
\end{abstract}

\maketitle

\section{Introduction}

In recent years, the study of gravitational theories in
asymptotically anti-de Sitter (AdS) spaces has been of great
interest due to the AdS/CFT correspondence
\cite{Juan,GKP,Witten,MAGOO}, a conjectured equivalence between at
least certain such ``bulk" string theories (which therefore
contain gravity) and non-gravitational dual theories.  In the case
of AdS, the non-gravitating dual theories are associated with
spacetimes that may be considered to form the boundary of the
asymptotically anti-de Sitter space. Similar so-called
gauge/gravity correspondences also arise for other systems (see
e.g. \cite{ISMY,HI,MR}) and involve bulk spacetimes with some of
the same features as anti-de Sitter space.

As one may expect, the notion of energy (and of other conserved
charges) is of significant use in understanding this
correspondence. For some time, it has been clear that the dual
field theories are closely associated with what is called the ``counter-term subtraction" definition of energy
\cite{skenderis,kraus,KS2,KS3,KS4,KS5,KS6,KS7,KS8} in the bulk.  Such ideas are well developed
for the case of anti-de Sitter space, and one might expect a
suitable generalization to apply to other contexts as well.
However, a number of other definitions of energy
\cite{asht1,asht2,ht,ad,gary,warner,t,KBL} have also been given for
bulk theories in AdS, and these are known to differ from the
counter-term subtraction definition.

In particular, these other definitions all assign zero energy to
pure AdS space, as is required if the charges are to form a
representation of the AdS group.  In contrast, in odd dimensions
the counter-term subtraction approach assigns a {\it non-zero}
value to AdS space which, moreover, depends on the choice of an
auxiliary structure: a conformal frame $\Omega$ at infinity.  This
feature is natural from the point of view of the dual gauge theory
(where it is associated with the conformal anomaly
\cite{skenderis,kraus}), but raises the question of the general
relationship between the counter-term subtraction energy and other
constructions.

A reasonable conjecture is that the difference between these
various notions of energy amounts to a ``constant offset'' which
might in general depend on the choice of auxiliary conformal frame
$\Omega$, but which in no way depends on the dynamical bulk
degrees of freedom.  If this were so, the difference would be a
constant over the phase space of the theory and all notions of
energy would generate the same action on observables via the
Poisson Bracket.  This conjecture is consistent with the
interpretation of the ``vacuum energy'' assigned to pure AdS  as
arising from the Casimir energy in the dual field theory.  It is
also suggested by numerous calculations (see e.g.,
\cite{skenderis,kraus,KS2,KS3,KS4,KS5,KS6,KS7,KS8,DGH}, and also
\cite{LS,GPP,MOTZ} for cases with slightly weaker asymptotic
conditions) of the value of the counter-term energy assigned to
particular families of spacetimes (e.g., the Schwarzschild-AdS
spacetimes) in a particular conformal frame and also by \cite{RT}.
Under appropriate asymptotically anti-de Sitter asymptotic
conditions, this conjecture was recently proven \cite{HIM} for all
solutions and in all conformal frames in $d=5$ bulk spacetime
dimensions.  Ref. \cite{HIM} also derives an explicit formula for
this difference as a function of the metric on the conformal
boundary defined by $\Omega$, and shows under their boundary
conditions that the definitions
\cite{asht1,asht2,ht,ad,gary,warner,t} also agree with a covariant
phase space definition based on techniques of \cite{ABR,wz,wi}.
Finally, since the appearance of the first version of the present
paper, \cite{KSnew} has extended such arguments to more general
asymptotically AdS boundary conditions.

Our purpose here is to demonstrate similar results in all
dimensions, and also for a much broader class of asymptotic
behaviors.  In fact, our arguments below will use only a few basic
features associated with the construction of counter-term charges.
We state most of the required properties in section \ref{cts}
below, but these properties follow immediately in cases where
counter-term subtraction is associated with the conformal boundary
of the spacetime manifold.  In addition, we will impose a simple
causality requirement in section \ref{main} which naturally occurs
whenever the conformal boundary has Lorentz signature.  Thus, our
results imply those of \cite{HIM,KSnew} and, in addition, apply
equally well to other contexts such as the domain-wall spacetimes
renormalized in \cite{BST,BBHV,CO} and to the cascading geometries
 renormalized in \cite{ABY} (and first studied in \cite{C1,C2,C3}).
Furthermore, if an appropriate set of counter-terms can be found,
they would also apply to the more general gauge/gravity dualities
described in \cite{ISMY}.

Our arguments will be based on general properties of the so-called
Peierls bracket \cite{peierls}, a manifestly covariant
construction which is equivalent to the Poisson bracket on the
space of observables (see \cite{gen} for extensions of the Peierls
bracket to algebras of gauge-dependent quantities and
\cite{Fred1,Fred2} for recent related work in quantum field
theory).  We begin by reviewing both the counter-term subtraction
definition of charge and the Peierls bracket in section
\ref{prelim}. This serves to set a number of conventions, and the
counter-term charge discussion provides an opportunity to comment
on subtle features associated with the choice of conformal frame
$\Omega$ used to define the charge associated with a particular
asymptotic symmetry $\xi$.  In particular, depending on the choice
of conformal frame, a given asymptotic symmetry need not act as a
strict symmetry on the collection of boundary fields used to
construct the counter-term charges.  Instead, it might act only as
a conformal symmetry.  However, in the special case of appropriate
asymptotically anti-de Sitter behavior, one may nevertheless show
\cite{skenderis,kraus,KS2,KS3,KS4,KS5,KS6,KS7,KS8} that the
difference between the charge evaluated on any two hypersurfaces
is determined entirely by the conformal frame $\Omega$ and is
independent of the bulk dynamics. Thus, even in this context the
counter-term definition remains useful.  We also take this
opportunity to generalize the construction to allow arbitrary
tensor and spinor boundary fields\footnote{The case of certain
scalar fields was considered in \cite{KS3,KS4,KS6,KS8}.  The
contribution of gauge fields to the divergence of the stress
tensor  was considered in \cite{gaugefields}.  In addition, we
understand that the corresponding conserved quantities are also
constructed in unpublished work by Kostas Skenderis, with results
similar to those presented below.}. Following this review, we give
our main argument in section \ref{main} and close with a brief
discussion of the results.

Since our arguments below will rely only on general properties of
the Peierls bracket, they are independent of the details of the
bulk dynamics. This is in sharp contrast to the results of
\cite{HIM,KSnew} which also compared various definitions of
energy, but which were based on a common technique involving
explicit expansion of the Einstein equations in a power series
around the boundary of an asymptotically anti-de Sitter space.
Our results here are correspondingly more general, but also much
less explicit.  We remind the reader that \cite{HIM} was able not
only to relate the counter-term energy to the covariant phase
space Hamiltonian, but also to show that the covariant phase space
Hamiltonian agrees with the constructions of Ashtekar et al based
on the electric part of the Weyl tensor \cite{asht1,asht2}, with
the Hamiltonian charge due to Henneaux and Teitelboim \cite{ht},
and finally with the spinor charge of \cite{gary,warner,t} (which
guarantees positivity). The
 Abbott and Deser construction \cite{ad} and its extensions \cite{SD2,SD3,OK} and the KBL construction \cite{KBL} (applied to AdS in \cite{KBLapp})
were not considered in \cite{HIM}.

\section{Preliminaries}
\label{prelim}

In this section, we review the two constructions central to our
analysis:  the counter-term subtraction definition of conserved
charges (section \ref{cts}) and the Peierls bracket (section
\ref{pb}) between observables.

\subsection{Counter-term Subtraction Charges}
\label{cts}

The setting for the counter-term subtraction construction of
conserved charges~\cite{skenderis,kraus} is to consider systems
associated with a certain sort of variational principle.  Now, in
general, such a principle specifies a class of variations with
respect to which one requires the associated action $S$ to be
stationary. Let us suppose that this is done by positing a space
of kinematically allowed histories ${\cal H}$ (``bulk variables'')
within which one is allowed to perform an arbitrary variation.
There will also be certain features (``boundary values") which are
identical for all histories in ${\cal H}$ and which are not to be
varied.  Thus, we in fact consider a family of actions $S$, each
with an associated space of histories ${\cal H}$, parameterized by
some set of allowed boundary values. Although typically discussed
in the context of the conformal completion of some spacetime, the
counter-term subtraction construction of conserved charges
generalizes naturally to a somewhat more abstract setting. We will
therefore find it useful to state the minimal axioms for this
construction.  The reader may readily verify that each axiom holds
when the boundary manifold $\partial M$ described below is the
conformal boundary of the spacetime $M$. Though our setting is in
principle more abstract, it is convenient to use the term
``boundary manifold'' and other such terms in our discussion.

The counter-term subtraction construction of conserved charges is
relevant when the following conditions hold:
\begin{enumerate}[1)]
\item The boundary values can be described by a set of tensor (and
perhaps spinor) ``boundary fields'' on an auxiliary manifold
$\partial M$ which is called the `boundary of the spacetime $M$.'
This will typically require the introduction of some auxiliary
structure, which we call $\Omega$, and which may include for
example a choice of conformal frame at infinity. The choice of
$\Omega$ is typically not unique, but is by definition a fixed
kinematical structure independent of the bulk state.  Given
$\Omega$, the boundary fields are determined by the bulk fields.

\item One of these boundary fields is a metric $h_{ab}$ on
$\partial M$ such that $(\partial M, h_{ab})$ is a globally
hyperbolic spacetime.

\item \label{diffeo} The action $S$ is diffeomorphism invariant in
the following sense:  Every diffeomorphism $\psi_\partial$ of the
boundary manifold $\partial M$ is (not uniquely) associated with a
diffeomorphism $\psi$ of the bulk spacetime which i) induces the
action of $\psi_\partial$ on the boundary fields through the map
that constructs boundary fields from bulk fields, ii) preserves
the auxiliary structure $\Omega$, iii) preserves the space ${\cal
H}$ of histories on which the action $S$ is defined, and iv) is
such that $S$ is invariant under the simultaneous action of $\psi$
on the bulk fields, $\psi_\partial$ on the boundary values, and
the corresponding transformation on the initial and/or final
boundary conditions appropriate to the action $S$. As a result,
the equations of motion are invariant under the action of $\psi$.
We refer to the vector fields generating $\psi$ and
$\psi_\partial$ as $\xi$ and $\xi_\partial$.  Note that only
diffeomorphisms $\psi$ for which $\psi_\partial$ acts as the
identity on $\partial M$ are gauge transformtations.

\item \label{finite} First functional derivatives of the action
$S$ with respect to the boundary fields are well-defined and
finite when evaluated on the space ${\cal S}$ of solutions to the
equations of motion. This is typically arranged by an appropriate
choice of ``counter-terms,'' leading to the name counter-term
subtraction method.
\end{enumerate}
As a particular example of this construction, one may consider
asymptotically anti-de Sitter spacetimes.  In this case, one takes
$\partial M$ to be the conformal boundary of $M$ defined by the
conformal frame $\Omega$.  The condition that $\psi$ in
requirement (\ref{diffeo}) above should preserve the conformal
frame $\Omega$ determines how $\psi_\partial$ is extended from
$\partial M$ to $M$, at least near $\partial M$.

In addition, we shall further assume:
\begin{enumerate}[5)]
\item \label{mult} Given $\xi$, $\xi_\partial$ as in
(\ref{diffeo}) and any smooth function $f$ on $M$, there is a
smooth function $f_\partial$ on $\partial M$ such that a the
action on bulk fields of a diffeomorphism along $f\xi$ induces the
action of a diffeomorphism on boundary fields along $f_\partial
\xi_\partial$.
\end{enumerate}
This latter condition clearly follows when the boundary $\partial
M$ is constructed by conformal completion of $M$, and will be
useful in our arguments below.

The above setting is somewhat analogous to consideration of a
field theory in the presence of non-dynamical background fields.
Here, however,  the role of the background fields is played only
by the boundary fields.  As a result, there is an important
difference: typically, one may vary background fields
independently of dynamical fields, such as when one constructs the
stress-energy tensor by varying a background metric for some field
theory in curved spacetime.  Clearly this is not possible here:
since the boundary fields are limiting values of the bulk fields,
any variation of the boundary fields necessarily requires a
corresponding variation of the bulk fields. This will lead to
certain subtleties which must be properly taken into account
below.

As a result, the current context will require more reliance on the
space of solutions (i.e., ``on-shell" techniques) than in the
usual background-field setting.  In particular, one makes heavy
use of the fact that,  when evaluated on the space of solutions,
variations which preserve both the boundary fields and appropriate
boundary conditions in the past and/or future will leave the
action invariant. It is this fact which allows property (4) above
to hold: as noted above, any variation of the boundary fields must
be accompanied by a variation of the bulk fields, and away from
the space of solutions the change in the action $S$ depends
non-trivially one the choice of bulk variation.  However, when
evaluated on-shell, the change in $S$ is independent of the choice
of bulk variation, so long as it satisfies appropriate boundary
conditions in the past and/or future. As a result, one may follow
\cite{skenderis,kraus,KS2,KS3,KS4,KS5,KS6,KS7,KS8,gaugefields} and
define the ``boundary stress tensor'' $\tau_{ab}$ as a function on
the space of solutions satisfying

\begin{equation}
\label{tau}
 \tau_{ab} \eps = -2 \frac{\delta S}{\delta
h^{ab}},
\end{equation}
where the functional derivative is computed holding all other
boundary fields constant and fixing appropraite boundary
conditions in the past and/or future.  Here we have used the
notation $\eps = \eps_{[a_1 a_2 \dots a_n]}$ for the natural
$n$-form associated with $h_{ab}$, identified with a density.

The definition (\ref{tau}) is sufficient when the metric is the
only non-trivial boundary field; i.e., in the context considered
by \cite{skenderis,kraus}.  In that context one may show that $\tau_{ab}$ is
covariantly conserved with respect to the metric $h_{ab}$ on
$\partial M$ by following the essential steps through which one
would derive covariant conservation of the stress-energy tensor
$T_{ab}$ in a curved spacetime.  We will describe this argument
below, but we also wish to consider the more general case in which
other boundary fields may be non-vanishing.  When the extra fields
are not scalars, this generalization will require us to introduce
a ``modified boundary  stress tensor'' with extra terms
representing contributions from these extra boundary fields.

To do so, let us introduce some complete set of bulk fields
$\Phi^I$ on $M$, where the $I$ ranges over an appropriate label
set to include components of vector and tensor (and perhaps
spinor) fields as well as scalars.  In particular, $\Phi^I$
includes the bulk metric (and any frame fields; see below).  We
also wish to pick out a complete set of boundary fields. However
it turns out that the tensor (or spinor) rank of these fields will
affect the detailed form of certain expressions below (including
the definition of the charges). As a result, it is convenient at
this stage to replace the boundary metric $h_{ab}$ with a set of
frame fields $e^{\phantom a A}_a$ satisfying
\begin{equation}
h_{ab} = \eta_{AB} e^{\phantom a A}_a e^{\phantom a B}_b
\end{equation}
for a fixed metric $\eta_{AB}$ (perhaps the Minkowski metric). The
introduction of the frame fields allows us to write all remaining
boundary fields without loss of generality in terms of a set of
{\it scalar} fields, e.g., a tensor field $X_{ab \dots c}$ is
encoded in the set $X_{AB \dots C} = X_{ab \dots c} e_A{}^a
e_B{}^b \cdots e_C{}^c$ of scalar fields. We denote the collection
of scalar fields on $\partial M$ by $\phi_\partial^i$. Thus, these
boundary scalars are just the `tangent space components' of any
remaining\footnote{The tangent space components of the frame
fields are, of course, trivial by definition.  These may be
included in the set $\phi_\partial^i$ for convenience of notation,
but only the set $\{\phi_\partial^i,e_A{}^a\}$ of boundary scalars
together with boundary frame fields forms a complete set of
boundary fields.} vector, tensor, or spinor boundary fields. We denote
the full set of such boundary scalar fields and the frame fields by
\ben
\Phi_\partial^I = (\phi^i, e_A{}^a) \, .
\een

Having replaced the boundary metric by a set of frame fields, it is natural
to introduce the ``modified boundary stress tensor''

\begin{equation}
\label{modtau} \T^{ab} \eps=    \frac{\delta
S}{\delta e^{\phantom a A}_b} e^{aA},
\end{equation}
where the functional derivative is computed holding fixed the
scalars $\phi^i_\partial$ (i.e., the tangent space components of
boundary fields).

More specifically, let us introduce the future and past boundaries
$\Sigma_\pm$ (perhaps at infinity) of our system in order to keep
track of all boundary terms.  We shall assume that, as is most
common, the action is chosen so that its functional derivatives
yield the equations of motion when boundary fields are held fixed
together with the fields\footnote{More generally, one might use an
action appropriate to fixing various derivatives of $\Phi^I$ at
$\Sigma_\pm$. It will be clear from the treatment below that our
results apply equally well to such cases.}
 $\Phi^I$ on $\Sigma_\pm$. Thus, a
general variation of the action may be written:
\begin{equation}
\label{varyS} \delta S =
 \int_M \frac{\delta S}{\delta \Phi^I}
\delta \Phi^I  + \int_{\partial M} \frac{\delta S}{\delta
\phi_\partial^i} \delta \phi^i_\partial + \int_{\partial M} \eps
\T^a_{\phantom a A} \delta e^{\phantom a A}_a +
\int_{\Sigma_{\pm}} \pi_I \delta \Phi^I, \end{equation} where
$\int_{\Sigma_\pm}$ includes integrals over both $\Sigma_+$ and
$\Sigma_-$ and we take the momenta $\pi_I$ to be defined by this
final term.  We are then interested in the value of
$\T^a_{\phantom a A}$ on the space of solutions.

In the case where the nontrivial boundary fields are just the
metric and some scalars on $\partial M$, the modified and original
boundary stress tensors agree; $\T_{ab} = \tau_{ab}$.  However, in
the presence of other non-trivial boundary fields, $\T^{ab}$
contains extra contributions from these fields.  As usual, we will
use the frame fields $e^{\phantom a A}_a$ and the inverse frames
to convert spacetime indices into tangent space indices (and vice
versa).  In particular, we will make use of $\T^{a}_{\phantom a
A}$, which is in fact a more fundamental quantity than $\T^{ab}$.

Now, in general, the modified boundary stress tensor $\T^A{}_a$
will fail to be covariantly conserved due to the presence of the
other background fields $\phi^i_\partial$. However, its covariant divergence takes a
simple and useful form. This may be demonstrated by considering
the simultaneous action of an arbitrary infinitesimal boundary
diffeomorphism $\psi_\partial$, which we take to be generated by
the vector field $\xi_\partial^a$, and the associated bulk
diffeomorphism $\psi$ generated by $\xi^a$.  By property
(\ref{diffeo}) above we then have

\begin{equation}
\label{v1}
 0 = \int_M \frac{\delta S}{\delta \Phi^I} \pounds_\xi
\Phi^I  + \int_{\partial M} \frac{\delta S}{\delta
\phi_\partial^i} \pounds_{\xi_\partial} \phi^i_\partial +
\int_{\partial M} \eps \T^a_{\phantom a A} \pounds_{\xi_\partial}
e^{\phantom a A}_a + \int_{\Sigma_{\pm}} \pi_I  \pounds_\xi
\Phi^I,
\end{equation}
If we evaluate (\ref{v1}) on the space of solutions (so that the
bulk equations of motion hold), then the first term vanishes.
Considering the second term,  the $\phi_\partial^i$ are scalars so
that we have $\pounds_{\xi_\partial} \phi^i_\partial =
\xi_\partial^a \nabla_a \phi^i_\partial$, where $\nabla$ is the
(torsion-free) covariant derivative on $\partial M$ compatible
with the metric $h_{ab}$.  Thus, this term is algebraic in
$\xi_\partial.$ Finally, turning to the third term, we have
\begin{equation}
\label{LieD} \pounds_{\xi_\partial} e^{\phantom a A}_a =
\xi_\partial^b \nabla_b e^{\phantom a A}_a + e_b^{\phantom a A}
\nabla_a\xi_{\partial}^b.\end{equation}

Thus, we may perform an integration by parts in the third term and
use the arbitrariness of $\xi_\partial^a$ (and, in particular, the
ability to set $\xi_\partial^a$ to zero in a neighborhood of
$\Sigma_\pm$) to conclude that the covariant divergence of
$\T_{ab}$ satisfies\footnote{Some readers may consider it more
elegant to introduce another derivative operator $D_a$ on
$\partial M$ satisfying $D_a^{} e^{\phantom a B}_b =0$. In this
case, $D_a \T^{ab}$ is given just by the scalar field term on the
right-hand side of (\ref{tdiv}).}
\begin{equation}
\label{tdiv} \nabla_a \T^{ab} = \sum_i
\frac{\delta S}{\delta \phi^i_\partial} \nabla^b \phi^i_\partial +
\T^a_{\phantom a A}
 \nabla^b e^{\phantom a A}_a.
 \end{equation}

We are now in a position to construct the counter-term charges and
demonstrate their conservation. To do so,  consider a particular
choice of boundary values and an infinitesimal diffeomorphism
$\psi_\partial$ corresponding to a symmetry of the boundary
values. We take $\psi_\partial$ to be generated by the vector
field $\xi_\partial$ and the associated bulk diffeomorphism $\psi$
to be generated by $\xi$. Hence $\xi_\partial$ Lie-derives the
boundary fields up to a local gauge transformation
\begin{equation}
\label{gaugerot}
\pounds_{\xi_\partial} e_a{}^A= R^A{}_B e_a{}^B,
\quad
\pounds_{\xi_\partial} \phi^i_\partial = \sum_j R^i{}_j \phi^j_\partial,
\end{equation}
where $R_{AB} = - R_{BA}$ and $R^i{}_j$ gives the action of the
associated frame rotation on the boundary scalars
$\phi^i_\partial$.  In fact, as we will see shortly, it is just as
easy to allow $(R^A{}_B,R^i{}_j)$ to define an arbitrary
infinitesimal transformation $\delta e_a{}^A= R^A{}_B e_a{}^B,
\delta  \phi^i_\partial = \sum_j R^i{}_j \phi^j_\partial$ under
which the action $S$ is locally invariant\footnote{By locally
invariant, we mean that
\begin{equation}
\label{locInv} \int_V \left( R^A{}_B e^B{}_a \frac{\delta}{\delta
e_a{}^A} + R^i{}_j \phi_\partial^j \frac{\delta}{ \delta
\phi^i_\partial} \right) S = 0
\end{equation} for any $V
\subset
\partial M$.  In particular, (\ref{locInv}) contains no boundary term on $\partial V$.}.

We call such a $\xi$ an ``asymptotic symmetry compatible with
$\Omega$."  One then defines the associated ``counter-term
subtraction charge:''

\begin{equation}
\label{qcounterdef} Q[\xi] = \int_C \T_{ab} \xi^a \, ds^b,
\end{equation}
where $C$ is a Cauchy surface of $\partial M$, and
$$ds^a = \eps^a{}_{b_1 \dots b_{n-1}}dx^{b_1} \cdots dx^{b_{n-1}}$$
is the induced integration element on $C$. We will refer to $C$ as
a `cut' of $\partial M$ in order to avoid confusion with Cauchy
surfaces in $M$. As an example of $Q[\xi]$ in the familiar anti-de
Sitter context, one might take the boundary metric to be the
Einstein static universe with all other boundary fields vanishing.
In this case, one could take $\xi$ to be an asymptotic time
translation and the associated $Q[\xi]$ would give the
counter-term subtraction definition of energy. Note also that we
have defined $Q[\xi]$ only when $\xi^a$ preserves any auxiliary
structure ($\Omega$) needed to define the boundary fields.
However, in typical examples (e.g., AdS) the result may be applied
much more generally: one need only find the boundary symmetry
$(\xi)_\partial$ associated with $\xi$ and then choose another
extension $\xi^{\prime}$ to the bulk which preserves $\Omega$ and
induces the same action $\xi_\partial^a$ on the boundary. One then
defines $Q[\xi] := Q[\xi']$.

We wish to prove that $Q[\xi]$ is independent of the choice of cut
$C$.  Let us therefore consider some region ${V} \subset \partial M$
such that the boundary ${\partial V}$ within $\partial M$ consists of
two cuts $C_1$ and $C_2$.  Let $Q_{C_1}[\xi]$ and $Q_{C_2}[\xi]$
denote the values of $Q[\xi]$ associated with the two cuts
respectively. Then we have

\begin{equation}
\label{diff} Q_{C_1}[\xi] - Q_{C_2}[\xi] = \int_{V} \eps
\nabla_a \left(\T^{ab} \xi_{\partial b} \right).
\end{equation}
But we  may use (\ref{LieD}) and (\ref{tdiv}) to express
(\ref{diff}) as
\begin{eqnarray}
\label{diff2} Q_{C_1}[\xi] - Q_{C_2}[\xi] &=& \int_{V} \sum_i
\frac{\delta S}{\delta \phi^i_\partial} \pounds_{\xi_\partial}
\phi^i_\partial + \int_{V} \eps \T^a_{\phantom a A}
\pounds_{\xi_\partial} e^{\phantom a
A}_a\nonumber\\
&=& \int_V \left( R^A{}_B e^B{}_a \frac{\delta}{\delta e_a{}^A} + \sum_{i,j}
R^i{}_j \phi_\partial^j \frac{\delta}{
\delta \phi^i_\partial} \right) S = 0 \,,
\end{eqnarray}
where in the second step we have used the fact that $\xi_\partial$
generates a symmetry of the  boundary fields up to a gauge rotation,
and where in the final step we have used the fact that $S$ is invariant under
such rotations.

Thus, for asymptotic symmetries $\xi$
compatible with $\Omega$, $Q[\xi]$ is indeed independent of the
cut $C$. Note that, as a result, we can weaken the framework to
require only that $C$ is homotopic to a Cauchy surface.
The result (\ref{diff2}) generalizes the construction of \cite{skenderis,kraus,KS2,KS3,KS4,KS5,KS6,KS7,KS8,gaugefields} to include arbitrary non-trivial (tensor and spinor) boundary fields.

\subsection{Conformal Boundary Killing Fields and Asymptotically anti-de
Sitter Boundary Conditions}

\label{conf}

In \cite{skenderis,kraus,KS2,KS3,KS4,KS5,KS6,KS7,KS8} it was shown
that many gravitational theories with asymptotically anti-de
Sitter asymptotic behavior satisfy requirements (1-5) of section
\ref{cts}. In addition,
\cite{skenderis,kraus,KS2,KS3,KS4,KS5,KS6,KS7,KS8} also
demonstrate another property associated with the conformal
invariance of the dual field theory (under the AdS/CFT
correspondence). Recall that conformal invariance requires the
trace of the stress-energy tensor to be zero. Now, if such a
quantum field theory is placed on a generic curved background the
trace of the stress tensor might be non-vanishing. This
trace---the ``anomaly'' ---is normally given by local curvature
terms of the background metric.  As a result, the AdS/CFT
correspondence suggests that the trace $\tau = h^{ab}\tau_{ab}$ of
the boundary stress tensor defined above should depend only on
$h_{ab}$ and, in particular, should be a constant on the space of
solutions ${\cal S}$. That this is the case was shown in
\cite{skenderis,kraus,KS2,KS3,KS4,KS5,KS6,KS7,KS8} for their
boundary conditions, under which $\tau^{ab}$ agrees with our
$\T_{ab}$. Indeed, when the metric on $\partial M$ is taken to be
the Einstein static universe (and certain other boundary fields
vanish), references
\cite{skenderis,kraus,KS2,KS3,KS4,KS5,KS6,KS7,KS8} show that
$\tau$ vanishes.

We may now follow
\cite{skenderis,kraus,KS2,KS3,KS4,KS5,KS6,KS7,KS8} and use this
observation to generalize the discussion of $Q[\xi]$ to the case
where $\xi$ is associated with a vector field $\xi_\partial$ on
$\partial M$ which is only a conformal killing field of $h_{ab}$.
Note that in cases where the boundary spacetime $(\partial
M,h_{ab})$ is just the conformal boundary of the bulk, any
asymptotic symmetry $\xi$ of the bulk should induce such a
conformal isometry $\xi_\partial$ of the boundary metric of
$\partial M$ so that this procedure will lead to a counter-term
charge associated with every conserved quantity that one expects
from the symmetries of the bulk system.

In particular, let us suppose that we have a conformal Killing
field with
\begin{equation}
\label{ckvf} \nabla_a \xi_{\partial b} + \nabla_b \xi_{\partial a}
= \pounds_{\xi_\partial} h_{ab} = 2k h_{ab},
\end{equation}
for some smooth function $k$ on $\partial M$, and that
\begin{equation}\label{phitrans}
\pounds_{\xi_\partial} \phi^i_\partial = \sum_j K^i{}_j \phi^j_\partial +  \sum_j R^i{}_j \phi^j_\partial \, .
\end{equation}
Here the coefficients $K^i{}_j$ encode the behavior of the $\phi_\partial^i$ under conformal transformations and the $R^i{}_j$ are as before in section \ref{cts}.
Equation~(\ref{ckvf}) implies that $\pounds_{\xi_\partial} e_a{}^A = k e_a{}^A + R^A{}_B e^B_a$.
We now simply repeat the above calculation to see how $Q_C[\xi]$
depends on the cut $C$.  Consider again equations (\ref{tdiv}) and (\ref{diff}), but
now use equation (\ref{ckvf}) to write the right-hand side in the form

\begin{eqnarray}
\label{diff2b} Q_{C_1}[\xi] - Q_{C_2}[\xi] &=& \int_{V}
\left( \eps \T^a{}_A \pounds_{\xi_\partial} e^A{}_a + \sum_i \frac{\delta S}{\delta
\phi^i_\partial} \pounds_{\xi_\partial} \phi^i_\partial  \right)
\cr
   &=&
\int_{V} \left(k\eps \T + \sum_{i,j} K_j{}^i \phi_\partial^j \frac{\delta S}{\delta \phi_\partial^i} \right)\\
&=& \int_V \left( k e^A{}_a \frac{\delta}{\delta e^A{}_a} +
\sum_{i,j} K^i{}_j \phi_\partial^j \frac{\delta}{\delta \phi^i_\partial} \right) S,
\end{eqnarray}
where we have defined $\T := \T^{a}_{\phantom a
A} e_{a}^{\phantom a A} = \T^{ab} h_{ab}$, and in the second line we have used the invariance of $S$ under frame rotations. Thus, assuming that
the integrand on the right side is
a function of the boundary fields alone and not of the particular solution under
consideration (as is the case under the asymptotic conditions
 of \cite{skenderis,kraus,KS2,KS3,KS4,KS5,KS6,KS7,KS8}), the change in $Q[\xi]$
 is a function only of the boundary fields and is otherwise
constant over the space of solutions ${\cal S}$.

\subsection{The Peierls bracket}
\label{pb}

Having reviewed (and generalized) the counter-term substraction
definition of charges, we now briefly review the other piece of
machinery we will need to derive our main result: the Peierls
bracket.

The Peierls bracket is an algebraic structure defined on
gauge-invariant functions on the space of solutions ${\cal S}$
associated with an action principle.  As shown in the original
work \cite{peierls}, this bracket is equivalent to the Poisson
bracket under the natural identification of the phase space with
the space of solutions.  One of the powerful features of the
Peierls bracket is that it is manifestly spacetime covariant.
Another is that it is defined directly for general gauge
invariants $A$ and $B$ whether or not $A$ and $B$ are associated
with some common time $t$. Furthermore, $A$ and $B$ need not be
local but can instead be extended over regions of space and time.

These features make the Peierls bracket ideal for studying the
boundary stress-tensor, which is well-defined only on the space of
solutions and is not a local function in the bulk
spacetime\footnote{For the same reasons, we expect the Peierls
bracket to be of use in studying other objects  which naturally
arise in the AdS/CFT correspondence.}.  As a result, it will be
straight-forward to give a Peierls version of a Noether argument
to show that the charges $Q[\xi]$ generate the appropriate
symmetries when $\xi_\partial$ is a boundary Killing field -- or,
more generally, a boundary conformal Killing field as discussed in
section \ref{conf}. Since this property is required of any charge
defined by Hamiltonian methods, it follows that such charges can
differ from $Q[\xi]$ only by a quantity with vanishing Peierls
bracket. But all such quantities can depend only on the boundary
fields and must otherwise be constants on the space of solutions
${\cal S}$.

The Peierls construction considers the effect on one gauge
invariant function (say, $B$) on the space of histories ${\cal H}$
when the action is deformed by a term proportional to the another
such function ($A$).  In particular, suppose that the dynamics is
determined by an action $S$.  One defines the advanced ($D^+_AB$)
and retarded ($D^-_AB$) effects of $A$ on $B$ by comparing the
original system with a new system defined by the action
$S_{\epsilon} = S + \epsilon A$, but associated with the same
space of histories. Here $\epsilon$ is a real parameter which will
soon be taken to be infinitesimal, and the new action is
associated with a new space ${\cal S}_\epsilon$ of deformed
solutions.

Under retarded (advanced) boundary conditions for which the
solutions $s \in {\cal S}$ and $s_{\epsilon} \in {\cal
S}_{\epsilon}$ coincide on $\Sigma_-$ ($\Sigma^+$) of the support
of $A$, the quantity $B_0 = B(s)$ computed using the undeformed
solution $s$ will in general differ from $B_{\epsilon}^\pm =
B(s_{\epsilon})$ computed using $s_{\epsilon}$ and retarded $(+)$
or advanced $(-)$ boundary conditions.  For small epsilon, the
difference between these quantities defines the retarded
(advanced) effect $D^-_AB$ ($D^+_AB$) of $A$ on $B$ through:
\begin{equation}
\label{effects} D^{\pm}_AB = \lim_{\epsilon \rightarrow 0}
\frac{1}{\epsilon}(B_{\epsilon}^\pm-B_0^{} ),
\end{equation}
which is a function of the unperturbed solution $s$.  Similarly,
one defines $D^\pm_BA$ by reversing the roles of $A$ and $B$
above. Since $A,B$ are gauge invariant, $D^\pm_BA$ is a
well-defined (and again gauge-invariant) function on the space
${\cal S}$ of solutions so long as both $A$ and $B$ are
first-differentiable on ${\cal H}$ (a requirement which may be
subtle when the spacetime supports of $A$ and $B$ extend to
$\Sigma_+$ or $\Sigma_-$).

The Peierls
bracket \cite{peierls} is then defined to be the difference of
the advanced and retarded effects:
\begin{equation}
\label{Peierls} \{ A,B \} = D^+_AB - D^-_AB.
\end{equation}
One may show that (\ref{Peierls}) depends only on the restriction
of $A,B$ to the space of solutions ${\cal S}$, so that
(\ref{Peierls}) defines an algebra of functions on ${\cal S}$, as
desired.

The fact that this agrees with the Poisson bracket (supplemented
by the equations of motion) was shown in \cite{peierls}, and
generalizes the familiar result that the commutator function for a free
scalar field is given by the difference between the advanced and
retarded Green's functions.  In fact, it is enlightening to write
the Peierls bracket more generally in terms of such Green's
functions.  To do so, we again make use of our complete set of (bulk) fields
$\phi^i$ (which include the metric and components of bulk tensor and
spinor fields) and the associated advanced and retarded Green's
functions $G_{IJ}^\pm(x,x')$.  Note that we have
\begin{equation}
\label{GFs} D^+_AB = \int dx \ dx' \frac{\delta B}{\delta
\Phi^I(x)} G^+_{IJ}(x,x') \frac{\delta A}{\delta \Phi^J(x')} =
 \int dx \ dx' \frac{\delta B}{\delta \Phi^J(x')} G^-_{JI}(x',x)
\frac{\delta A}{\delta \Phi^J(x)} = D^-_BA.
\end{equation}
Thus, the Peierls bracket may also be written in the manifestly
anti-symmetric form
\begin{equation}
\label{Peierls2} \{ A,B \} = D^-_BA-D^-_AB.
\end{equation}
The expressions (\ref{GFs}) in terms of $G^\pm_{IJ}(x,x')$ are
also useful in order to verify that the Peierls bracket defines a
Lie-Poisson algebra.  In particular, the derivation property
$\{A,BC\}= \{ A,B \} C + \{A,C\} B$ follows immediately from the
Leibnitz rule for functional derivatives.  The Jacobi identity
also follows by a straightforward calculation, making use of the
fact that functional derivatives of the action commute (see e.g.,
\cite{Bryce1,Bryce2}). If one desires, one may use related Green's
function techniques to extend the Peierls bracket to a Lie algebra
of gauge dependent quantities \cite{gen}.

\section{Main Argument}

\label{main}

We now use the Peierls bracket to show that the counter-term
subtraction charges $Q[\xi]$ generate the appropriate symmetries
when $\xi_\partial$ is a boundary Killing field, or, more
generally, a boundary conformal Killing field under the conditions
of section \ref{conf}. Since this property is required of any
charge defined by Hamiltonian methods, it follows that such
charges can differ from $Q[\xi]$ only by a quantity with vanishing
Peierls bracket.  But any such quantity can be built only from
auxiliary structures and must otherwise be constant on the space
of solutions ${\cal S}$. As in section \ref{prelim}, we first
address asymptotic symmetries $\xi$ compatible with $\Omega$ using
the features (1-5) of the counter-term subtraction setting as
described in section \ref{cts}, and then proceed to the case where
$\xi$ does not preserve $\Omega$ so that the associated
$\xi_\partial$ acts only as a conformal Killing field on the
boundary.

\subsection{Asymptotic Symmetries Compatible with $\Omega$}

The essential point of the argument is that the Peierls bracket
allows a simple derivation of Noether's theorem.  We will be able
to proceed when there is a pair of smooth functions
$(f,f_\partial)$ on $(M,\partial M)$ satisfying requirement (5) of
section \ref{cts} as well as
\begin{itemize}
\item  $f=0$ in a neighborhood of the past boundary $\Sigma_-$.
\item $f_\partial =0$ to the past of some cut $C_0$ of $\partial
M$.
\item $f=1$ in a neighborhood of the future boundary
$\Sigma_+$. \item $f_\partial=1$ to the future of some cut $C_1$
of $\partial M$.
\end{itemize}
This is the simple causality requirement mentioned in the
introduction.  It is naturally satisfied whenever $\partial M$ may
be considered as a boundary of $M$ and is of Lorentz signature. In
that case we may simply take $f_\partial$ to be defined by limits
of $f$ on $\partial M$.

Let us now consider any asymptotic symmetry $\xi$ compatible with
$\Omega$ and the associated boundary isometry $\xi_\partial$.
Under the action of this symmetry, the bulk and boundary fields
transform as
\begin{equation}
\label{SymAct} \delta \Phi^I = \pounds_{\xi}\Phi^I, \quad \delta
e_a{}^A = \pounds_{\xi_\partial} e_a{}^A= R^A{}_B e_a{}^B, \quad
{\rm and} \quad \delta \phi^i_\partial = \pounds_{\xi_\partial}
\phi^i_\partial = \sum_j R^i{}_j \phi^j_\partial,
\end{equation}
where $(R^A{}_B, R^i{}_j)$ provide an appropriate frame rotation
of the boundary fields.

The key point of our argument is to construct a new transformation
$\Delta_{f,\xi}$ on the space of fields such that the associated
first order change $\Delta_{f,\xi} S$ in the action generates the
asymptotic symmetric $\xi$.  We will see that the correct choice
is given by
$\Delta_{f,\xi} \Phi^I := (\pounds_{f
\xi} - f\pounds_{\xi} )\Phi^I$. An important property of this
definition is that the change $\Delta_{f,\xi} \Phi^I$ is {\it
algebraic} in $\Phi^I$; i.e., we need not take spacetime
derivatives of the fields $\Phi^I$ in order to compute
$\Delta_{f,\xi} \Phi^I$. Furthermore, $\Delta_{f,\xi} \Phi^I$ is
proportional to $\nabla_a f$, and thus vanishes in a neighborhood
of $\Sigma_+$ and $\Sigma_-$.  This property guarantees that
$\Delta_{f,\xi} S$ is differentiable on the space ${\cal H}$ of
histories associated with the action $S$.  In particular, solutions to the equations of motion resulting
from the deformed action  $S + \epsilon \Delta_{f,\xi} S$ are stationary points of  $S + \epsilon \Delta_{f,\xi} S$ under {\it all}
variations $\delta \Phi^I$ which preserve the boundary fields
$\phi_\partial^i, e_a{}^A$ (up to gauge rotations) and vanish on $\Sigma_\pm$; all boundary terms vanish under arbitrary such variations.

As an additional consequence of the above, we see that (on-shell)
the quantity $\Delta_{f,\xi} S$ is gauge-invariant:  Since the
action $S$ is gauge-invariant, the quantity $\Delta_{f,\xi} S$ can
acquire gauge dependence only through $f, \xi$.  However, the
above observation and (\ref{varyS}) imply that on-shell
$\Delta_{f,\xi} S$ depends only on $f_\partial, \xi_\partial$.
Since gauge transformations have trivial action on $\partial M$,
we conclude that $\Delta_{f,\xi} S$ is gauge-invariant on-shell.
Thus, we may take the Peierls bracket of $\Delta_{f,\xi} S$ with
any other on-shell observable $A$.

To do so, let us note that if $s \in {\cal S}$ is a stationary
point of the original action $S$ with bulk fields $\Phi^I$ and
boundary fields $\phi^i_\partial$, $e_a{}^A$, then to first order
in $\epsilon$ we see that $s_1 = (1-\epsilon\Delta_{f,\xi}) s$ is
a stationary point of $S + \epsilon \Delta_{f,\xi_i} S$, since to
first order this modified action is just $S(\Phi^I + \epsilon
\Delta_{f,\xi_i}\Phi^I)$; i.e., we see that to first order the
bulk fields are merely shifted by $- \epsilon \Delta_{f,\xi}$.
Since $\xi$ is an asymptotic symmetry compatible with $\Omega$,
property (\ref{diffeo}) of section \ref{cts} states that the
boundary fields defined by $s_1$ are also shifted by $- \epsilon
\Delta_{f,\xi}$ relative to those of $s$.

Of course, we desire solutions to the modified equations of motion
whose boundary values give the {\it original} boundary fields of
$s$. However, this can be arranged by making use of another
symmetry. Note that because $\xi$ is an asymptotic symmetry, we
may use (\ref{gaugerot}) to compute the induced action of
$\Delta_{f,\xi}$ on boundary fields as follows:
\begin{eqnarray}
\label{bound1} \Delta_{f,\xi} \phi_\partial^i &=&
(\pounds_{f_\partial \xi_\partial} - f_\partial
\pounds_{\xi_\partial}) \phi^i_\partial = \pounds_{f_\partial \xi_\partial}
\phi^i - f_\partial R^i{}_j \phi^j_\partial, \cr \Delta_{f,\xi} e_a{}^A &=&
(\pounds_{f_\partial \xi_\partial} - f_\partial
\pounds_{\xi_\partial} ) e_a{}^A = \pounds_{f_\partial
\xi_\partial} e_a{}^A - f_\partial R^A{}_B e_a{}^A.
\end{eqnarray}
Thus, the shift of the boundary fields is just given by the a
diffeomorphism along the vector field $-f_\partial \xi_\partial$
and a compensating frame rotation $(R^i{}_j, R^A{}_B)$. In fact,
as will shortly be important, the shift $\Delta_{f,\xi}
\phi_\partial^i$ of the boundary scalars vanishes (and the shift
$\Delta_{f,\xi} e_a{}^A$ of the boundary frame fields simplifies
dramatically) using (\ref{gaugerot}), but for the moment the form
(\ref{bound1}) is more useful.  To see why, recall that the
equations of motion are invariant under both diffeomorphisms and
frame rotations. As a result, if $R$ is a frame rotation on the bulk fields which induces
the rotation $(R^i{}_j, R^A{}_B)$ on the boundary, then
\begin{equation}
\label{s2} s_2 = (1 + \epsilon \pounds_{f \xi} - \epsilon f R) s_1
= (1 + \epsilon f \pounds_{\xi} - \epsilon f R) s
\end{equation}
 with bulk fields
\begin{equation}
\Phi^I|_{s_2} = \Phi^I - \epsilon (\Delta_{f,\xi} -
\pounds_{f\xi}) \Phi^I - \epsilon fR^I{}_J \Phi^J= \Phi^I +
\epsilon f \pounds_\xi \Phi^I - \epsilon fR^I{}_J \Phi^J
\end{equation}
induces the original boundary fields (by (\ref{bound1}))
\begin{equation}
\label{s2bf} \phi^i_\partial|_{s_2} = \phi^i_\partial|_s, \quad
e_a{}^A|_{s_2} = e_a{}^A|_s,
\end{equation}
and again solves the equations of motion that follow from
$S(\Phi^I + \epsilon \Delta_{f,\xi_i} \Phi^I)$.

We may use this observation to compute the advanced and retarded
changes $D^\pm_{\Delta_{f,\xi}S}A$ of any gauge invariant quantity
$A$.  Let us first consider the retarded change, and let us
evaluate this change on a solution $s$ as above.  We seek a
solution $s^-_\epsilon$ of the perturbed equations of motion which
agrees with $s$ on $\Sigma_-$.  Since the infinitesimal
transformation $f(\pounds_\xi-R)$ vanishes on $\Sigma_-$, we see
that we may set $s_\epsilon^- = s_2$ as defined (\ref{s2}) above;
i.e. $s_\epsilon^- = (1+ \epsilon f[\pounds_\xi-R]) s$. Thus, the
retarded effect on $A$ is just $D^-_{\Delta_{f,\xi}S}A = f
\pounds_\xi A,$ where we have used the fact that $A$ must be
invariant under local frame rotations.

To compute the advanced effect, we seek a solution $s^+_\epsilon$
of the perturbed equations of motion which agrees with $s$ on
$\Sigma_+$. Consider the history $s^+_\epsilon = (1 -
\epsilon[\pounds_\xi-R])s^-_\epsilon = (1 + (f-1)\epsilon[
\pounds_\xi-R])s$. Since this differs from $s^-_\epsilon$ by the
action of a symmetry compatible with $\Omega$, it again solves the
equations of motion (to first order in $\epsilon$) and induces the
required boundary fields (\ref{s2bf}).  In addition, since $f=1$
on $\Sigma_+$, we see that $s^+_\epsilon$ and $s$ agree on there.
Thus, we may use $s^+_\epsilon$ to compute the advanced change in
any gauge invariant $A$:
\begin{equation}
D^+_{\Delta_{f,\xi}S}A = (f-1) \pounds_\xi A.
\end{equation}
Finally, we arrive at the Peierls bracket
\begin{equation}
\label{PBfinal} \{\Delta_{f,\xi}S   ,  A  \} =
D^+_{\Delta_{f,\xi}S } A - D^-_{\Delta_{f,\xi}S } A = -
\pounds_\xi A.
\end{equation}
Thus, $-\Delta_{f,\xi}S$ generates a diffeomorphism along the
asymptotic symmetry $\xi$ as desired\footnote{The form of
$\Delta_{f,\xi}S$ is similar to the Hamilton-Jacobi definition of
energy proposed in \cite{RS} in the context of asymptotically flat
space. As a result, a similar argument might also be used to
demonstrate equivalence of such a construction with Hamiltonian
methods in that context.}.

Our task is now to relate $\Delta_{f,\xi} S$ to $Q[\xi]$. But this
is straightforward.  From (\ref{varyS}), we have
\begin{equation}
\label{DSvar} \Delta_{f,\xi} S =
 \int_M \frac{\delta S}{\delta \Phi^I} \Delta_{f,\xi} \Phi^I
 + \int_{\partial M} \frac{\delta
S}{\delta \phi_\partial^i} \Delta_{f,\xi} \phi^i_\partial +
\int_{\partial M} \eps \T^a_{\phantom a A}\Delta_{f,\xi}
 e^{\phantom a A}_a +
\int_{\Sigma_{\pm}} \pi_I \Delta_{f,\xi} \Phi^I.
\end{equation}
However, $\Delta_{f,\xi} \Phi^I$ vanishes on $\Sigma_\pm$ and on
the boundary fields we may use (\ref{LieD}) to find:
\begin{eqnarray}
\label{bound2} \Delta_{f,\xi} \phi_\partial^i &=&
(\pounds_{f_\partial \xi_\partial} -
f_\partial\pounds_{\xi_\partial} ) \phi^i_\partial = 0,
 \cr
\Delta_{f,\xi} e_a^A &=& (\pounds_{f_\partial \xi_\partial} -
f_\partial\pounds_{\xi_\partial} )  e_a^A =e_b^A \xi_\partial^b
\nabla_a f_\partial .
\end{eqnarray}
Thus, on-shell, only the term containing $\T^a_{\phantom a A}
\nabla_a f_\partial$ contributes to (\ref{DSvar}).

Furthermore, since $f_\partial$ is constant both to the past of
$C_0$ and to the future of $C_1$, we may replace the integral over
$\partial M$ with an integral over the region ${V}$ between $C_0$
and $C_1$. Thus, (\ref{DSvar}) takes the form

\begin{eqnarray}
\Delta_{f,\xi} S =
  &=& -   \int_{{V}}  \eps \T^{ab}
(\xi_\partial)_b \nabla_a f \cr &=& - \int_{C_1} \T_{ab} \xi^b \,
ds^a +  \int_{V}  \eps f \nabla_a \left( \T^{ab}
(\xi_\partial)_b \right)
  \cr &=& - \int_{C_1} \T_{ab} \xi^b \, ds^a
+ \int_V f  
\left( R^A{}_B e^B{}_a \frac{\delta}{\delta e_a{}^A} + \sum_{i,j}
R^i{}_j \phi_\partial^j \frac{\delta}{
\delta \phi^i_\partial} \right) S 
\cr &=& - Q_{C_1}[\xi].
\end{eqnarray}
In the second line, we have used that $\xi$ is an
asymptotic symmetry (see eqs.(\ref{diff}, \ref{diff2})), 
and that the action is invariant under frame rotations.
In passing from the first to the second line we have used the fact
that $f_\partial=0$ on $C_0$.

Thus,  $-\Delta_{f,\xi} S$ agrees (on-shell) with the charge
$Q[\xi]$ evaluated on the cut $C_1$.  By the arguments of section
\ref{cts}, this equality also holds on any other cut of $\partial
M$. Consequently, since by eq.~(\ref{PBfinal}) the variation
$\Delta_{f,\xi} S$ generates the action of the infinitesimal
symmetry $\xi$ on observables, it follows that the same must be
true for the counter-term charges. Thus,
\begin{equation}
\label{result} \{ Q[\xi], A \} = \pounds_\xi A,
\end{equation}
as desired.

\subsection{Asymptotic Symmetries not compatible with $\Omega$}

In fact, we may apply a similar argument to the case described in
section \ref{conf}, where an asymptotic symmetry $\xi$ is {\it
not} compatible with $\Omega$ and is thus associated with a
boundary vector field $\xi_\partial$ which is only a conformal
Killing field of the chosen boundary fields, see~(\ref{ckvf},
\ref{phitrans}).  As such cases are not addressed by the axioms
stated in section \ref{cts}, we state the corresponding
requirements here.  We will derive our results when the following
additional conditions hold:

\begin{enumerate}[6)]
\item Under the action of a diffeomorphism along $f \xi$ on a
history (i.e., $h \rightarrow (1 + \epsilon \pounds_{f\xi} h$) the
boundary fields transform with additional conformal weights:
\begin{equation}
\label{bfs} \delta_{ \pounds_{f\xi}} \phi^i_\partial =
\pounds_{f_\partial \xi_\partial} \phi^i_\partial  + f_\partial
K^i{}_j \phi^j , \quad \delta_{ \pounds_{f\xi}} e_a{}^A =
\pounds_{f_\partial \xi_\partial} e_a{}^A  + f_\partial K^A{}_B
e_a{}^B.
\end{equation}
\end{enumerate}

\begin{enumerate}[7)]
\item
Furthermore, since $\xi_\partial$ is a conformal symmetry of the
boundary fields and $K^i{}_j,K^A{}_B$ are the associated conformal
weights, the right-hand side of (\ref{bfs}) becomes just a frame
rotation when $f=1$.
\end{enumerate}
The reader may readily check that requirements (6) and (7) above
are fulfilled by the usual setting for counter-term subtraction
schemes in asymptotically AdS spaces.

As a result of requirements (6) and (7), the histories
$s^\pm_\epsilon$ identified in section III.A above (see, e.g.,
(\ref{s2}))  again have boundary values identical to those
($\phi^i_\partial, e_a{}^A$) of $s$. Thus we may proceed exactly
as above to again conclude:

\begin{equation}
\label{PBfinal2} \{\Delta_{f,\xi}S   ,  A  \} =
D^+_{\Delta_{f,\xi}S } A - D^-_{\Delta_{f,\xi}S } A = -
\pounds_\xi A.
\end{equation}

Furthermore, using property (6) we find that, on-shell, we may
calculate $\Delta_{f,\xi}S$:
\begin{eqnarray}
\Delta_{f,\xi}S &=&
 -   \int_{{V}}  \eps \T_{ab}
(\xi)^b_\partial \nabla^a f  - \int_{{V}} f\left( k e^A{}_a
\frac{\delta}{\delta e^A{}_a} + \sum_{i,j} K{}^i{}_j
\phi_\partial^j \frac{\delta}{\delta \phi^i_\partial} \right) S
\cr
 &=& - \int_{C_1} \T_{ab} \xi^b_\partial \, ds^a +  \int_{{V}}  f \nabla_a \left( \T^{ab} \xi_b \right) -
 \int_{{V}} f\left( k e^A{}_a \frac{\delta}{\delta e^A{}_a} +
\sum_{i,j} K{}^i{}_j \phi_\partial^j \frac{\delta}{\delta
\phi^i_\partial} \right) S
  \cr &=& - \int_{C_1} \T_{ab} \xi^b_\partial \, ds^a  +  \int_{{V}}f \frac{\delta S}{\delta
\phi_\partial^i} \pounds_{\xi_\partial} \phi^i_\partial +
\int_{{V}} \eps f \T^a_A \pounds_{\xi_\partial} e^{\phantom a A}_a
\cr && -\int_{{V}} f\left( k e^A{}_a \frac{\delta}{\delta e^A{}_a}
+ \sum_{i,j} K{}^i{}_j \phi_\partial^j \frac{\delta}{\delta
\phi^i_\partial} \right) S \cr &=& - Q_{C_1}[\xi],
\end{eqnarray}
where in the last step we have again used property (7). Finally,
since we saw in section \ref{conf} that $Q_{C}[\xi]$ depends on
the cut $C$ only through a term that is constant on ${\cal S}$, it
follows that we have~(\ref{result}) for any cut $C$.  Thus,  even
when $(\xi)_\partial^a$ is only a conformal symmetry of the
boundary, $Q_C[\xi]$ can differ from any Hamiltonian generator of
the symmetry $\xi$ only through a (possibly cut-dependent) term
which is a function {\it only} of the boundary fields and which is
otherwise constant over the space ${\cal S}$ of solutions.

\section{Discussion}
We have used general arguments based on the Peierls bracket to
compare the counter-term subtraction charges $Q[\xi]$ of
\cite{skenderis,kraus,KS2,KS3,KS4,KS5,KS6,KS7,KS8} with any
Hamiltonian charges $H[\xi]$ when $\xi$ is a diffeomorphism which
generates a symmetry of an appropriate system. Specifically, when
$\xi$ induces a symmetry $\xi_\partial$ of the boundary fields, we
have shown that $Q[\xi]$ generates the bulk symmetry associated
with $\xi$ via the Peierls bracket.  As a result, it can differ
from $H[\xi]$ only by a term determined entirely by the boundary
fields and which is otherwise constant on the space of solutions.
Furthermore, since both $Q[\xi]$ and $H[\xi]$ are conserved, this
difference is also independent of the cut of infinity on which it
is evaluated.

Our results generalize a conclusion of \cite{HIM}, which was in
turn suggested by a number of more specific calculations (e.g.
\cite{skenderis,kraus,LS,GPP,MOTZ,RT}). Ref. \cite{HIM} showed via
direct calculation that $Q[\xi] - H[\xi]$ was a function of
boundary fields alone in $d=5$ spacetime dimensions and under a
particular set of asymptotic conditions; indeed, \cite{HIM} gives
an explicit formula for this difference. Ref. \cite{HIM} was also
able to show that $H[\xi]$ agrees with a definition of energy in
that context due to Ashtekar et al \cite{asht1,asht2}. However,
from the results of the present paper and the convention that the
Hamiltonian charges $H[\xi]$ vanish in AdS space, we may conclude
generally that $H[\xi] = Q[\xi] - Q[\xi](AdS)$,  where
$Q[\xi]({AdS})$ is the result obtained by evaluating the
counter-term charge in pure AdS space\footnote{After the
appearance of the first version of this work, the same result was
also derived in \cite{KSnew} under fairly general asymptotically
AdS boundary conditions.}. The present results may also be applied
in non-conformal versions of gauge-gravity duality (such as those
described in, e.g., \cite{ISMY}) if an appropriate set of
counterterms can be identified to implement requirements (1-5) of
section \ref{prelim}.  In particular, due to \cite{ABY} we may
apply them directly to certain spacetimes dual to cascading gauge
theories and, due to \cite{BST,BBHV,CO}, to domain-wall
spacetimes.

In addition, the work above generalizes the counter-term procedure
for constructing conserved charges to the case in which arbitrary
(tensor and spinor) non-trivial boundary fields may be present in
addition to the boundary metric. The result is simply the
replacement of the boundary stress tensor with the ``modified
boundary stress tensor'' $\T^{ab}$ of equation (\ref{modtau}),
which contains extra terms arising from any non-trivial boundary
fields which are not scalars.  This modified boundary stress
tensor is not covariantly conserved, and even boundary scalar
fields contribute to its divergence. Nevertheless, the form of
$\nabla_a \T^{ab}$ allows one to show that $Q[\xi]$ is in fact
conserved.  Furthermore, the Peierls bracket argument again shows
that $H[\xi] - Q[\xi]$ is constant on the space of solutions.

We also addressed a special case which arises when the bulk theory
is dual to a conformal theory, as in the original anti-de Sitter
context. In such cases, the counter-term action changes under a
conformal transformation, but only by a function of the boundary
fields which is otherwise constant on the space ${\cal S}$ of
solutions. As a result, one may consider the case of a vector
field $\xi_\partial$ which acts only as a conformal symmetry on
the boundary.  The result is again that $Q[\xi]$ generates the
action of the bulk symmetry along $\xi$ via the Peierls bracket
and thus that $Q[\xi]$ can differ from any Hamiltonian charge
$H[\xi]$ only by a term built from the boundary fields (and which
is otherwise constant on ${\cal S}$).  However, in this case the
term can depend (through a solution-independent term) on the cut
$C$ of the boundary spacetime on which it is
evaluated\footnote{Note that this dependence vanishes for the
special case of asymptotically AdS spaces when the boundary metric
is chosen to be the Einstein static universe and all other
boundary fields vanish, since in that case $\tau=\tau^{ab}h_{ab} =
0$ \cite{skenderis,kraus,KS2,KS3,KS4,KS5,KS6,KS7,KS8}.}.

Recall that when $\partial M$ is determined through conformal compactification (as in the asymptotically anti-de Sitter context), any asymptotic symmetry induces a conformal Killing field on the boundary.  Thus, in this case one may work
with a fixed conformal structure $\Omega$ and still construct all
conserved quantities via the counter-term subtraction method.  Furthermore, Hamiltonian generators which vanish on AdS
space itself are given for all asymptotic symmetries $\xi$ by

\begin{equation}
H[\xi] = Q_C[\xi] - Q_C[\xi](AdS),
\end{equation}
where we have once again subtracted off the value $Q_C[\xi](AdS)$ of the counter-term charge evaluated on a corresponding cut $C$ of $\partial M$ in pure
anti de-Sitter space.
As a result, both $H[\xi]$ and $Q_C[\xi]$ are consistent with the
covariant phase space methods of \cite{wz}, which controls only variations
of the Hamiltonian on the space of solutions.

\begin{acknowledgments} D.M. would like to thank Vijay Balasubramanian and Per
Kraus  for useful discussions about counter-term charges, and
Chris Beetle, Rafael Sorkin, and other members of the Perimeter
Institute seminar audience for useful discussions on the Peierls
bracket. S.H. and D.M. were supported in part by NSF grant
PHY0354978 and by funds from the University of California. S.H.
was also supported in part by DOE grant DE-FG02-91ER40618. D.M.
was also supported in part by funds from the Perimeter Institute
of Theoretical Physics.  A.I. was supported in part by NSF grant
PHY 00-90138 to the University of Chicago.
\end{acknowledgments}

\end{document}